\documentclass[prl,aps,twocolumn,showpacs,superscriptaddress]{revtex4} %or {revtex4-1}

    \usepackage{color} % dvips allows for colors
    \usepackage[colorlinks]{hyperref} %% hyperlinks
    \usepackage{graphicx,amssymb,amsbsy}
    \usepackage{ifthen}
    \usepackage[percent]{overpic}
    \newcommand{\wwwcb}[1]{
                  {\tt \href{http://ChaosBook.org#1}
              {ChaosBook.org#1}}}

\graphicspath{{figs/}}  %% directories with graphics files

\newcommand{\Lg}{\ensuremath{T}}   % 2014-04-04 prettier Lie algebra generator
\newcommand{\matrixRep}{\ensuremath{{D}}}  %  matrix rep of a group element
\newcommand{\rf}     [1] {~\cite{#1}}
\newcommand{\refref} [1] {ref.~\cite{#1}}

\newcommand{\refeq}  [1] {(\ref{#1})}
\newcommand{\reffig} [1] {fig.~\ref{#1}}
\newcommand{\refFig} [1] {Fig.~\ref{#1}}

\newcommand{\beq}{\begin{equation}}
\newcommand{\continue}{\nonumber \\ }

\newcommand{\eeq}{\end{equation}}
\newcommand{\ee}[1] {\label{#1} \end{equation}}
\newcommand{\bea}{\begin{eqnarray}}

\newcommand{\eea}{\end{eqnarray}}

\newcommand{\etc}{{\em etc.}}       % etcetera in italics
\newcommand{\ie}{{i.e.}}            % APS

\newcommand{\etal}{{\em et al.}}    % etal in italics, APS too

\newcommand{\statesp}{state space}

\newcommand{\dmn}{-dimensional}  %  n-dimensional

\newcommand{\pS}{{\cal M}}          % symbol for phase space
\newcommand{\zeit}{\ensuremath{\tau}}  %time variable
\newcommand\period[1]{{\ensuremath{\zeit_{#1}}}}         %continuous cycle period

\newcommand{\reals}{\mathbb{R}}

\renewcommand\Im{\ensuremath{{\rm Im}\,}}
\renewcommand\Re{\ensuremath{{\rm Re}\,}}

\newcommand{\po}{periodic orbit}

\newcommand{\rpo}{rela\-ti\-ve periodic orbit}
\newcommand{\Rpo}{Rela\-ti\-ve periodic orbit}
\newcommand{\eqv}{equi\-lib\-rium}

\newcommand{\reqv}{rela\-ti\-ve equi\-lib\-rium}
\newcommand{\reqva}{rela\-ti\-ve equi\-lib\-ria}
\newcommand{\bseq}{\begin{subequations}}
\newcommand{\eseq}{\end{subequations}}

\newcommand{\NSe}{Navier-Stokes equations}

\newcommand{\twomode}{two-mode}
\newcommand{\slicePlane}{slice hyperplane}

\newcommand{\reducedsp}{reduced state space}

\newcommand{\slice}{slice}

\newcommand{\mslices}{method of slices}

\newcommand{\template}{template} % {slice-fixing point} % {reference state}
\newcommand{\sliceBord}{slice border}

\newcommand{\pSRed}{\ensuremath{\hat{\cal M}}} % reduced state space Jan 2012
\newcommand{\sspRed}{\ensuremath{\hat{\ssp}}}    % reduced state space point Jan 2012
\newcommand{\velRed}{\ensuremath{\hat{\vel}}}    % ES reduced state space velocity Jan 2012

\newcommand{\slicep}{{\ensuremath{\sspRed'}}}   % slice-fixing point Jan 2012
\newcommand{\sliceTan}[1]{\ensuremath{t'_{#1}}}    % group orbit tangent at slice-fixing
\newcommand{\groupTan}{\ensuremath{t}}    % group orbit tangent
\newcommand{\phaseVel}{phase velocity}      % pipe slicing
\newcommand{\Un}[1]{\ensuremath{\textrm{U}(#1)}}         % in DasBuch
\newcommand{\SOn}[1]{\ensuremath{\textrm{SO}(#1)}}         % in DasBuch

\newcommand{\KS}{Kuramoto-Siva\-shin\-sky}
\newcommand{\KSe}{Kuramoto-Siva\-shin\-sky equation}

\newcommand{\REQV}[2]{\ensuremath{TW_{#1#2}}} % #1 is + or -
\newcommand{\eigExp}[1][]{
     \ifthenelse{\equal{#1}{}}{\ensuremath{\lambda}}{\ensuremath{\lambda^{(#1)}}}}
\newcommand{\eigRe}[1][]{
     \ifthenelse{\equal{#1}{}}{\ensuremath{\mu}}{\ensuremath{\mu^{(#1)}}}}
\newcommand{\eigIm}[1][]{
     \ifthenelse{\equal{#1}{}}{\ensuremath{\omega}}{\ensuremath{\omega^{(#1)}}}}

\newcommand{\vel}{\ensuremath{v}}   % state space velocity
\newcommand{\ssp}{\ensuremath{a}}                % state space point
\newcommand{\cssp}{\ensuremath{\tilde{u}}}                % Complex state space point
\newcommand{\csspRed}{\ensuremath{\hat{u}}}      % Symmetry reduced complex state space point
\newcommand{\fFslice}{first Fourier mode slice}

\newcommand{\braket}[2]
		   {\langle{#1}\vphantom{#2}|\vphantom{#1}{#2}\rangle}

\usepackage{amsmath}

\begin{document}
\title{
Reduction of SO(2) symmetry for
spatially extended dynamical systems
}
\author{Nazmi Burak Budanur}
\author{Predrag Cvitanovi\'c}
\affiliation{
                Center for Nonlinear Science, School of Physics,
                Georgia Institute of Technology,
                Atlanta, GA 30332-0430
               }
\author{Ruslan L. Davidchack}
\affiliation{Department of Mathematics,
	      University of Leicester, Leicester LE1 7RH, UK}
\author{Evangelos Siminos}
\affiliation{
Max-Planck Institute for the Physics of Complex Systems,
	  N\"{o}thnitzer Str. 38, D-01187 Dresden, Germany
            }

\date{\today}

        \begin{abstract}
Spatially extended systems, such as channel or pipe flows, are often equivariant
under continuous symmetry transformations, with each state of the flow having an
infinite number of equivalent solutions obtained from it by a translation or a
rotation. This multitude of equivalent solutions tends to obscure the
dynamics of turbulence. Here we describe the  `{\fFslice}', a very simple, easy
to implement reduction of SO(2) symmetry. While the method exhibits rapid
variations in phase velocity whenever the magnitude of the first Fourier mode is
nearly vanishing, these near singularities can be regularized by a time-scaling
transformation. We show that after application of the method, hitherto unseen
global structures, for example Kuramoto-Sivashinsky relative periodic orbits and
unstable manifolds of travelling waves, are uncovered.
        \end{abstract}


\maketitle

Mounting evidence that exact coherent structures play a key role in
shaping turbulent flows\rf{science04} necessitates developing new tools
for elucidating how these structures are interrelated\rf{GHCW07}.
Unravelling these interrelations for flows which admit continuous
symmetries requires special care. A solution to a problem in classical or
quantum mechanics starts with the classification of problem's symmetries,
followed by a choice of a {basis} invariant under these symmetries. For
example, one formulates the two-body problem in three {Cartesian}
coordinates, but when it comes to solving it, polar coordinates, with the
phase along the symmetry direction as an explicit coordinate, are
preferable. While a classification of problem's symmetries might be
relatively straightforward, for high-dimensional nonlinear systems
(fluids, nonlinear optical media, reaction-diffusion systems, \etc) a
good choice of a symmetry-invariant frame is not as easy as transforming
to polar coordinates. In this letter we describe the `{\fFslice}', a
simple method for reducing \Un{1}\ or \SOn{2}\ symmetry that has been
tested on and works well for systems of dimensions ranging from 4 (for
the `{\twomode} system'\rf{BuBoCvSi14}) to $10^5$ (for fluid
dynamics\rf{WiShCv14}).

The applications we have in mind are to {solutions
of} spatially extended systems, such as \NSe\ for a velocity field $u$
{on a spatially periodic domain}, where one starts the symmetry analysis
by rewriting the equations in a Fourier basis,
\beq
  u(x,\zeit)=\sum_{k=-\infty}^{+\infty} \cssp_k (\zeit)\, e^{ i q_k x }
\,,
\ee{eq:ksexp}
where $\cssp_k=x_k+i\,y_k = |\cssp_k| e^{i \phi_k}$,
$q_k = 2 \pi k / L$, $L$ is the domain size, $x$ is the spatial
coordinate and $\tau$ is time.
Thus a nonlinear PDE is converted to an infinite tower of ODEs.
In computations this \statesp\ is truncated to
$2m$ real dimensions
        \footnote{
The \statesp\ becomes $(2m\!+\!1)$-dimensional if one includes the $0$th 
Fourier mode $\cssp_0$, which is an invariant of translations \refeq 
{eq:U1}. By Galilean invariance of the \KSe, which we use for the 
demonstrations in this letter, $\cssp_0$ is a conserved quantity. 
Conventionally the average $\int_{0}^{L}\!u\,dx$ is set equal to zero, \ie, 
$\cssp_0=0$.
},
$\ssp = (x_1, y_1, x_2, y_2, \ldots, x_m,
y_m)^\mathsf{T}$. If the system has a translational symmetry, the
complex Fourier modes \refeq{eq:ksexp} form a continuous family of states
(a group orbit), equivalent under spatial translations $u(x, \zeit)
\rightarrow u(x + \delta x, \zeit)$, and related by \Un{1} rotations \beq
\cssp_k \rightarrow \cssp_k e^{i k \theta} \,,\quad \theta = 2 \pi \delta
x / L \,. \ee{eq:U1} In other words, the formulation contains a redundant
degree of freedom. Keeping such redundant degrees of freedom, as we shall
illustrate here with the \KS\ example, obscures the dynamics.

In this letter we shall assume that for a generic `turbulent' state $u(x,
\zeit)$ the first Fourier mode never exactly vanishes, and define the
symmetry-reduced Fourier modes $\csspRed_k$ by fixing the phase of the
first Fourier mode,
\beq
	\csspRed_k (\zeit) =
      e^{- i k \phi_1 (\zeit)} \cssp_k (\zeit)
	\, ,  \label{eq:FixFirstMode}
\eeq
The symmetry reduced Fourier modes $\csspRed_k$ are invariant under the
symmetry transformation \refeq{eq:U1} by construction. A phase-fixing
transformation of this kind is very natural; the earliest example known
to authors is the reduction of the $S^1$ symmetry of the complex
Ginzburg-Landau equation by Luce\rf{Luce95}. When applied to
spatiotemporally chaotic dynamics, however, the phase-fixing
transformation \refeq{eq:FixFirstMode} introduces what appear to be
discontinuities in the flow.
In this letter we show that a reexamination of the \mslices%
\rf{rowley_reconstruction_2000,BeTh04,SiCvi10,FrCv11,atlas12,ACHKW11}
leads to a regularization of such apparent singularities by means of a
rescaled `slice time'. This representation (from here on referred to as
the `{\fFslice}') reveals relations among important coherent structures
of the flow, such as \reqva\ and \rpo s, known to play an
important role in shaping the state space of turbulent
flows\rf{science04}. Here, for simplicity, we illustrate the {\fFslice}
by applying it to the dynamics of {\KSe} in
one spatial dimension. As shown in \refref{WiShCv14}, the method is
equally easily
incorporated into spectral codes for $3D$ fluid flows in periodic domains,
with no need for any further generalization.

Consider a first-order flow $\dot{\ssp} = \vel (\ssp)$ on \statesp\ $
\ssp \in \pS$ obtained from the $m$-mode truncation of the Fourier
expansion \refeq{eq:ksexp}. Here the velocity function $\vel (\ssp)
= (\dot{x}_1, \dot{y}_1, \dot{x}_2, \dot{y}_2 \ldots, \dot{x}_m, \dot{y}_m)^\mathsf{T}$
is the Fourier transform of the right side of the PDE
for field $u(x, \zeit)$. Translational symmetry in the configuration space
implies that the dynamics satisfies the equivariance condition
\beq
  \vel (\ssp) = \matrixRep(\theta)^{-1} \vel ( \matrixRep(\theta) \ssp )
\,,
\ee{eqvarcond}
where
\beq
	\matrixRep(\theta) = \mathrm{diag}\left[\,R(\theta),\, R(2 \theta),\, \ldots,\, R (m \theta)\,\right]\,,
\ee{mmodeLieEl}
is a block-diagonal $[2m\!\times\!2m]$ matrix representation
of the $\SOn{2}$ action and $R (k \theta)$
is the $[2\!\times\!2]$ rotation matrix acting on the $k$-th Fourier
mode. The generator of rotations is also a block-diagonal matrix, with
$[2\!\times\!2]$ infinitesimal generators of infinitesimal rotations
$\Lg_k$ along its diagonal,
\beq
	 R(k \theta) =  \begin{pmatrix}
			\cos k\theta & -\sin k\theta   \\
			\sin k\theta  & ~\cos k\theta
                \end{pmatrix}
\,,\quad
	 \Lg_k =  \begin{pmatrix}
			 0 & -k  \\
			k & 0
                \end{pmatrix}
\,.
\ee{mmodeLg}
For visualization
purposes we find it more convenient to work in the real \SOn{2}\ representation
rather than the complex $\Un{1}$ formulation \refeq{eq:U1}.
The group orbit $\pS_\ssp$ of a \statesp\ point \ssp\ is the set of all
points reachable from \ssp\ by symmetry transformations,
$
 \pS_\ssp = \{ \matrixRep(\theta ) \, \ssp\ | \, \theta \in [0, 2 \pi ) \}
\,.
$
In the \mslices, one constructs a `\slice', a submanifold $\pSRed \subset
\pS$ that cuts each group orbit in an open neighborhood once and only
once. The dynamics is then separated into the `shape-changing' dynamics
$\sspRed (\zeit) \in \pSRed $ within this submanifold, and a
symmetry coordinate parametrized by the group parameter $\theta (\zeit) $
(a `moving frame'\rf{CartanMF,FelsOlver98,Mansfield10}) that reconstructs
the original dynamics $\ssp (\zeit) \in \pS$ by the group action $\ssp
(\zeit) = \matrixRep(\theta (\zeit))\,\sspRed (\zeit) $. For the $\SOn{2}$
case at hand, a one-parameter family of transformations, $\pSRed $ has one
dimension less than $\pS$.

\begin{figure} %dasbuch/book/FigSrc/inkscape
\begin{center}
 \setlength{\unitlength}{0.4\textwidth}
 %% \unitlength = units used in the Picture Environment
 \begin{picture}(1,0.79592472)%
    \put(0,0){\includegraphics[width=\unitlength]{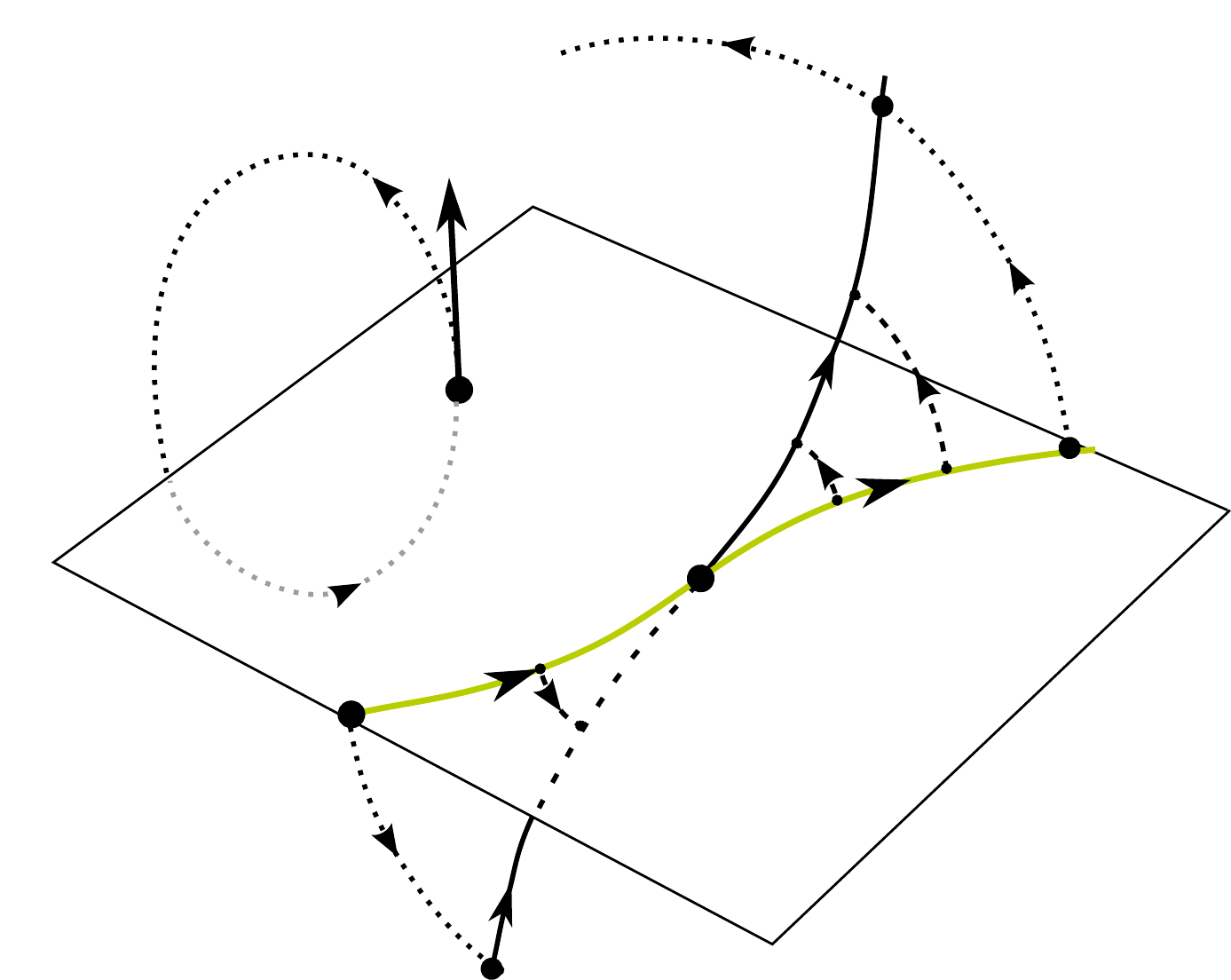}}%
    \put(0.12712463,0.34706651){\color[rgb]{0,0,0}\rotatebox{-30.34758661}{\makebox(0,0)[lb]{\smash{$\pSRed$}}}}%
    \put(0.74886166,0.72963928){\color[rgb]{0,0,0}\rotatebox{0.0313674}{\makebox(0,0)[lb]{\smash{$\ssp(\zeit)$}}}}%
    \put(0.90484057,0.46048552){\color[rgb]{0,0,0}\rotatebox{0.0313674}{\makebox(0,0)[lb]{\smash{$\sspRed(\zeit)$}}}}%
%    \put(0.21383593,0.75368385){\color[rgb]{0,0,0}\rotatebox{0.0313674}{\makebox(0,0)[lb]{\smash{$\matrixRep(\theta(\zeit))\ssp(\zeit)$}}}}%
    \put(0.40198025,0.45548326){\color[rgb]{0,0,0}\rotatebox{0.0313674}{\makebox(0,0)[lb]{\smash{$\slicep$}}}}%
    \put(0.38795968,0.6477644){\color[rgb]{0,0,0}\rotatebox{0.0313674}{\makebox(0,0)[lb]{\smash{$\sliceTan{}$}}}}%
    \put(0.60974703,0.28008178){\color[rgb]{0,0,0}\rotatebox{0.0313674}{\makebox(0,0)[lb]{\smash{$\sspRed(0)$}}}}%
 \end{picture}%
\end{center}
\caption{\label{f-ReducTraj1}
(Color online) The \slicePlane\ \pSRed, which passes through the
{\template} point $\slicep$ and is normal to its group tangent
$\sliceTan{}$, intersects all group orbits (dotted lines) in an open
neighborhood of $\slicep$.  The full \statesp\ trajectory $\ssp(\tau)$
(solid black line) and the \reducedsp\ trajectory $\sspRed(\zeit)$ (solid
green line) belong to the same group orbit $\pS_{\ssp(\zeit)}$ and are
equivalent up to a `moving frame' rotation by phase
$\theta(\zeit)$. Adapted from \wwwcb.
}%
\end{figure}

There is a great deal of freedom in how one constructs a \slice; in general one
can pick any `moving frame'.
Computationally easiest way to construct a local
\slice\ is by considering a hyperplane of points \sspRed\ defined by
\beq
0 = \braket{\sspRed}{ \sliceTan{} }
    \,,\quad\mbox{where}\quad
\braket{b}{c} = \sum_{\ell=1}^{2m} b_\ell c_\ell
\,,
\ee{slicecond}
is sketched in \reffig{f-ReducTraj1}. Here, $\sliceTan{} = \Lg \slicep$ is
the group tangent (the direction of translations) evaluated at a
reference \statesp\ point $\slicep$, or
`\template'\rf{rowley_reconstruction_2000}.
The template is assumed not to lie in an invariant subspace, \ie,
$\matrixRep(\theta)\slicep \neq \slicep$ for all $\matrixRep(\theta) \neq 1$.

The dynamics within this
\slicePlane\ and the reconstruction equation for the phase parameter are
given by
\bea
\velRed(\sspRed) &=& \vel(\sspRed)
   -\dot{\theta}(\sspRed) \, \groupTan(\sspRed)
                \label{e:velRed} \, , \\
\dot{\theta}(\sspRed) &=& {\braket{\vel(\sspRed)}{\sliceTan{}}}/
               {\braket{\groupTan(\sspRed ) }{\sliceTan{} } }
                \label{veltheta}
\,,
\eea
with $\groupTan(\sspRed) = \Lg \sspRed$ the group tangent evaluated at
the symmetry-reduced \statesp\ point $\sspRed$. Eq.~\refeq{e:velRed} says
that the full \statesp\ velocity $\vel(\sspRed)$ is the sum of the
in-slice velocity $\velRed(\sspRed)$ and the transverse velocity
$\dot{\theta}(\sspRed) \, \groupTan(\sspRed)$ along the group tangent,
and \refeq{veltheta} is the reconstruction equation whose integral tracks
the trajectory in the full \statesp\ (for a derivation and further references,
see \refref{DasBuch}).

The {\phaseVel} \refeq{veltheta} becomes singular
for $\sspRed^*$ such that
$\groupTan(\sspRed^*)$ lies in the \slice,
\beq
\braket{\groupTan(\sspRed^*)}{\sliceTan{}} = 0
\,,
\ee{ChartBordCond}
or for $\sspRed^*$ in an invariant subspace, where
$\groupTan(\sspRed^*)=0$. The $(d\!-\!2)$\dmn\ hyperplane of such points
$\sspRed^*$ forms the `\sliceBord', beyond which the \slice\ does not
apply. Both the \slicePlane\ and its border depend on the choice of
\template\ $\slicep$, with the resulting `chart' in general valid only in
some neighborhood of $\slicep$. For a turbulent flow, symmetry reduction
might require construction of a set of such local overlapping
charts\rf{ACHKW11,atlas12}. However, as we now show for $\SOn{2}$,
a simple choice of \template\ may suffice to avoid all \sliceBord\
singularities in regions of dynamical interest.

We define the `{\fFslice}' by choosing
\beq
 \slicep = (1,0,0,0,...)
\,,\quad
 \sliceTan{} = (0,1,0,0,...)
\,.
\ee{e:slicep}
The \slice\ determined by this \template\ is the $(d\!-\!1)$\dmn\ half-hyperplane
\bea
 \hat{x}_1 & \geq & 0 \,,\quad  \hat{y}_1 \,=\, 0\,, \continue
 \hat{x}_k, \hat{y}_k & \in & \reals\,,\quad \mbox{for all~~} k > 1
\,.
\label{e:sspRed_1}
\eea
The condition $\hat{y}_1 = 0$ follows from the \slice\ condition
\refeq{slicecond}, whereas $\hat{x}_1  \geq  0$ ensures a single intersection
for every group orbit. This choice of \slice\
corresponds to \refeq{eq:FixFirstMode}, fixing the phase of the first Fourier mode.
Now that we have the equations
\refeq{e:velRed} and \refeq{veltheta} for the dynamics in the \slicePlane\ we see
why this phase fixing transformation can run into singularities:
\sliceBord\ \refeq{ChartBordCond} for the \template\
\refeq{e:slicep} is located at $\hat{x}_1 = 0$, and the denominator
of \refeq{veltheta} approaches zero as the trajectory approaches the \sliceBord.
We regularize this singularity
 by defining the in-\slice\ time as
$d\hat{\zeit} = d{\zeit} / \hat{x}_1$, and rewriting \refeq{e:velRed} and
\refeq{veltheta} as
\bea
{d \sspRed}/{d \hat{\zeit}}          &=& \hat{x}_1 \vel(\sspRed)
    - \dot{y}_1(\sspRed) \,  \groupTan(\sspRed) \, ,
\label{e:velredscaledtime} \\
{d \theta (\sspRed)}/{d \hat{\zeit}} &=& \dot{y}_1(\sspRed)
\label{velthetascaledtime}
\,.
\eea
The {\phaseVel} \refeq{velthetascaledtime} is
now  the non-singular, full \statesp\ velocity component $\dot{y}_1$ orthogonal
to the \slice, and the full \statesp\ time is the
integral
\beq
\zeit(\hat{\zeit}) = \int_0^{\hat{\zeit}} \!\!\! d\hat{\zeit}'\, \hat{x}_1(\hat{\zeit}')
\,.
   \label{timerescaledtime}
\eeq
For example, the full \statesp\ period $\period{p}=
\period{p}(\hat{\zeit}_p)$ of a \rpo\ $\ssp(\period{p})=g_p\,\ssp(0)$
is the integral \refeq{timerescaledtime} over one period $\hat{\zeit}_p$
in the \slice.
\renewcommand{\ssp}{a}             % state space point
\renewcommand{\sspRed}{\ensuremath{\hat{\ssp}}}    % reduced state space point
\renewcommand{\slicep}{\ensuremath{\hat{\ssp}'}}   % slice-fixing point

%%%%%%%%%%%%%%%%%%%%%%%%%%%%%%%%%%%%%%%%%%%%%%%%%%%%%%%%%%%%%%%%%%%%%%%
\begin{figure}[tbp]
\centering
\begin{overpic}[width=0.23\textwidth]{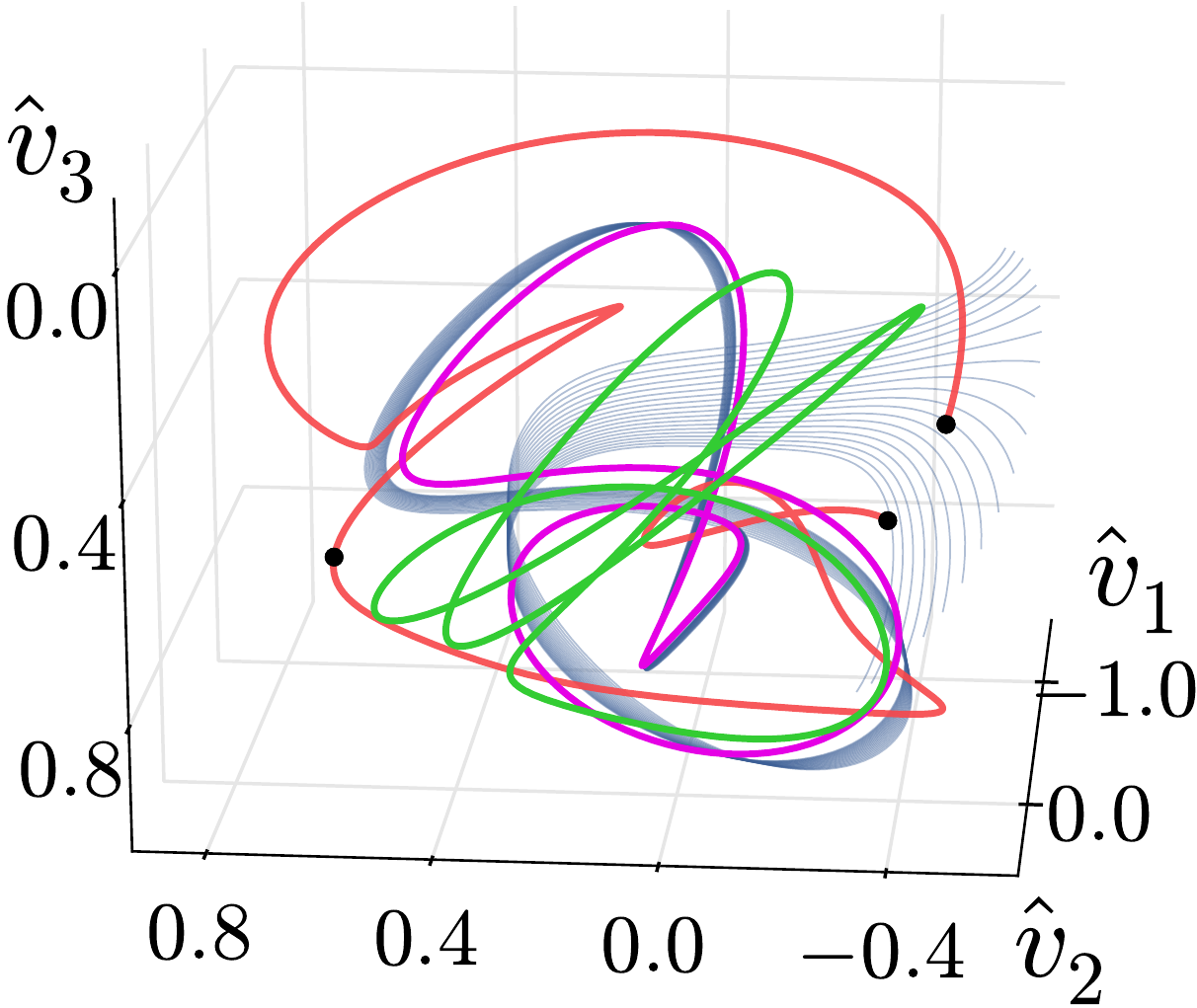}
                             \put (0,0) {(a)}
\end{overpic}\,
\begin{overpic}[width=0.23\textwidth]{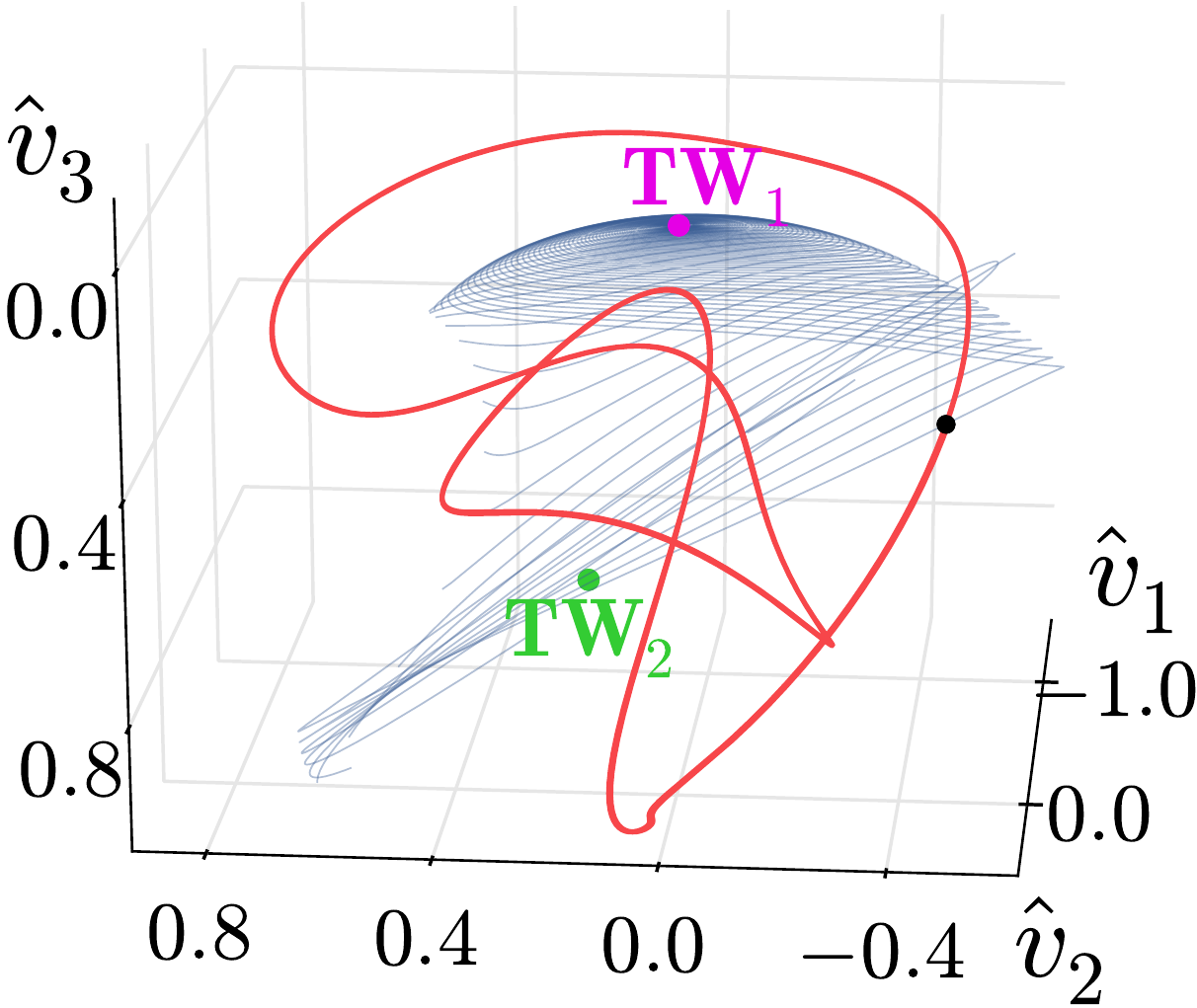}
                             \put (0,0) {(b)}
\end{overpic}
\caption[]{
(Color online)
\KS\ system
(a)  in the full \statesp: Unstable manifold of the \reqv\ $\REQV{}{1}$ (blue) traced out by
integrating nearby points given by \refeq{e-TWman_init2}: 2 repeats of
the $\period{p} = 33.5010$ \rpo\ (red), with instants $\zeit = 0,
\period{p}, 2\period{p}$ marked by black dots; group orbits (which are
also the time orbits) of the $\REQV{}{1}$ (magenta) and $\REQV{}{2}$
(green).
(b) In the symmetry reduced \statesp\
$\REQV{}{1}$ and $\REQV{}{2}$ orbits are reduced to single
points, the unstable manifold is a smooth 2D surface,
and the \rpo\ closes after a single period.
}
\label{f:ks}
\end{figure}
%%%%%%%%%%%%%%%%%%%%%%%%%%%%%%%%%%%%%%%%%%%%%%%%%%%%%%%%%%%%%%%%%%%%%%

We illustrate the utility of {\fFslice} by applying it to the \KS\
system on a periodic domain in one spatial dimension,
\[
  u_t = -{\textstyle\frac{1}{2}}(u^2)_x-u_{xx}-u_{xxxx} \, ,
\]
a model PDE extensively studied as it exhibits spatiotemporal
chaos\rf{Holmes96}.
The \reqv\ (traveling wave) $\REQV{}{i}$ and \rpo\ solutions
that we use in this example are described in \refref{SCD07},
where the domain size has been set to $L=22$, large enough
to exhibit complex spatiotemporal dynamics.
In terms of complex Fourier modes \refeq{eq:ksexp} the \KSe\ takes form:
\beq
\dot{\cssp}_k = ( q_k^2 - q_k^4 )\, \cssp_k
    - i \frac{q_k}{2} \sum_{m=-\infty}^{+\infty} \cssp_m \cssp_{k-m} \,.
\ee{expan}
In the real
representation $\cssp_k=x_k+i\,y_k$, \KSe\ is equivariant under \SOn{2}
rotations \refeq{mmodeLieEl}. We have adapted the ETDRK4
method\rf{cox02jcomp,ks05com} for numerical integration of the symmetry
reduced equations \refeq{e:velredscaledtime}, where we set $\cssp_0=0$
and truncate the expansion \refeq{expan} to $m=15$ Fourier modes, so the
\statesp\ is 30\dmn, $a = (x_1, y_1, x_2, y_2, ..., x_{15}, y_{15})^\mathsf{T}$.

As an illustration of symmetry reduction, we trace out a segment of the unstable
manifold of the \reqv\ $\REQV{}{1}$
by integrating $n$ trajectories for time $\zeit$, with initial conditions
$\sspRed_1, \cdots, \sspRed_n$ on the tangent vector $\hat{e}_1$,
\beq
  \sspRed_\ell = \sspRed_{{\REQV{}{1}}} + \epsilon \, e^{\ell\delta} \hat{e}_1
\,,\enspace
  \mbox{where}
\enspace
    \delta = 2 \pi \eigRe[1] /n \,\eigIm[1] \, .
\label{e-TWman_init2}
\eeq
Here $\sspRed_{{\REQV{}{1}}}$ is the point of intersection of the $\REQV{}{1}$
orbit with the \slicePlane, $n$ we set to 20,
integration time we set to $\zeit= 115$,
$\epsilon$  is a small parameter that
we set to $10^{-6}$, and  $\hat{e}_1 = \Re\,\hat{V}_1 / |\Re\,\hat{V}_1|$.
The unstable manifold of $\REQV{}{1}$ is
four-dimensional, with $\hat{V}_1$, $\hat{V}_2$
the expanding complex stability eigenvectors of
$\REQV{}{1}$ with eigenvalues $\eigExp[j] = \eigRe[j] \pm i\,\eigIm[j]$.
Here we present the two-dimensional submanifold associated with the most
expanding complex eigenvector $\hat{V}_1$.
\refFig{f:ks} shows the \statesp\ projections of the unstable
manifold of $\REQV{}{1}$, along with the $\period{p}=33.5010$ \rpo\ and the
\reqv\  $\REQV{}{2}$.
The coordinate axes are projections $(v_1, v_2, v_3)$ onto three
orthonormal vectors $(\hat{e}_1, \hat{e}_2, \hat{e}_3)$
constructed from $\Re\,\hat{V}_1$, $\Im\,\hat{V}_1$ and $\Re\,\hat{V}_2$
via Gram-Schmidt orthogonalization.
It is clear from \reffig{f:ks}\,(a) that without the
symmetry reduction,
the $\REQV{}{1}$ unstable manifold is dominated by the drifts along
its group orbit. In the symmetry reduced \statesp\ $\pSRed$,
\reffig{f:ks}\,(b), the dynamically important, group-action transverse
part of the unstable manifold of $\REQV{}{1}$ is revealed. While the
drifts along the symmetry direction complicate the \rpo\ in
\reffig{f:ks}\,(a), the same orbit closes onto itself after one repeat
within the \slicePlane, \reffig{f:ks}\,(b).
Likewise, $\REQV{}{2}$, which is topologically a circle but appears
convoluted in the projection of \reffig{f:ks}(a),
is reduced to a single \eqv\ point.
The stage is now set for a construction of symbolic dynamics for the flow
by means of Poincar\'e sections and return maps\rf{lanCvit07}.

%%%%%%%%%%%%%%%%%%%%%%%%%%%%%%%%%%%%%%%%%%%%%%%%%%%%%%%%%%%%%%%%%%%%%%%
\begin{figure}[h]
    \centering
        \begin{tabular}{c c c}
            \begin{overpic}[height=0.19\textheight]{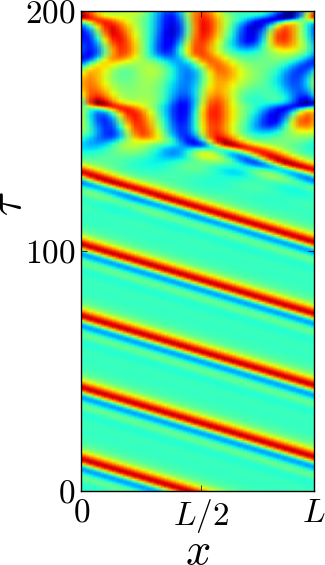}
                             \put (3,3) {(a)}
            \end{overpic}\,&
            \begin{overpic}[height=0.19\textheight]{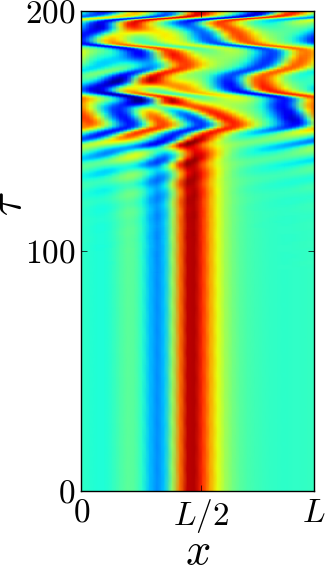}
                             \put (3,3) {(b)}
            \end{overpic}&
            \begin{overpic}[height=0.19\textheight]{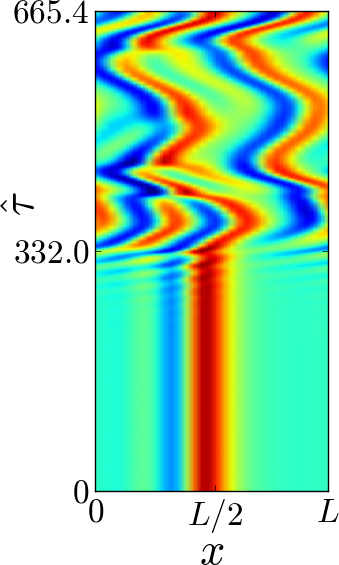}
                             \put (6,3) {(c)}
            \end{overpic}\\
            \begin{overpic}[height=0.19\textheight]{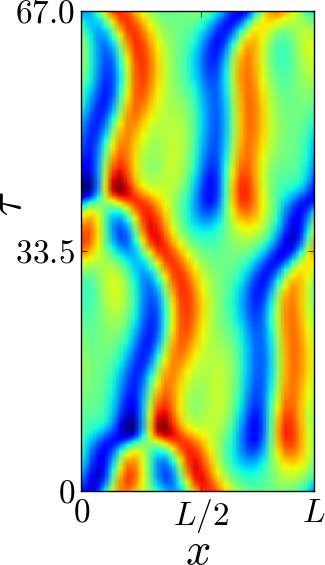}
                             \put (3,3) {(d)}
            \end{overpic}\,&
            \begin{overpic}[height=0.19\textheight]{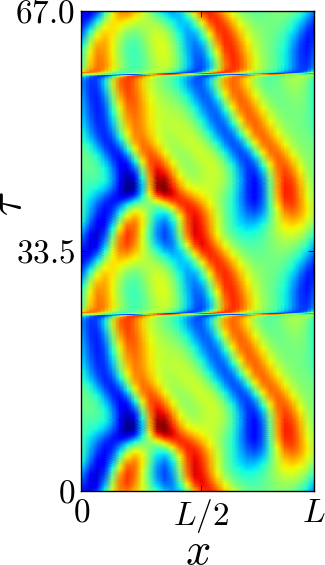}
                             \put (3,3) {(e)}
            \end{overpic}&
            \begin{overpic}[height=0.19\textheight]{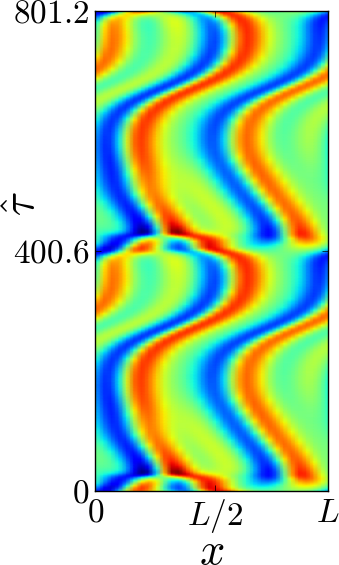}
                             \put (6,3) {(f)}
            \end{overpic}
        \end{tabular}
	\caption{
	(Color online)
	Traveling wave $\REQV{}{1}$ with {\phaseVel} $c = 0.737$
	in configuration space: (a) the full \statesp\ solution,
	(b) symmetry-reduced
	solution with respect to the lab time, and (c) symmetry-reduced solution with
	respect to the in-\slice\ time.
	\Rpo\  $\period{p} = 33.50$ in configuration space:
	(d) the full \statesp\ solution, (e) symmetry-reduced solution with respect
	to the lab time, and (f) symmetry-reduced solution with respect to the in-\slice\
	time.
	}
  \label{fig:f-ksconf}
\end{figure}
%%%%%%%%%%%%%%%%%%%%%%%%%%%%%%%%%%%%%%%%%%%%%%%%%%%%%%%%%%%%%%%%%%%%%%

The solutions of \KS\ system are conventionally visualized in the
configuration space, as time evolution of color-coded value of the
function $u(x,t)$. \refFig{fig:f-ksconf}\,(b,e) illustrates that a
\reqv\ and a \rpo\ become an \eqv\ and a \po\ after symmetry reduction.
\refFig{fig:f-ksconf}\,(a,b,c) shows that a numerical trajectory
eventually diverges from the unstable \reqv\ and falls onto the strange
attractor.
The sharp shifts along $x$ direction in \reffig{fig:f-ksconf}\,(e)
correspond to the time intervals where trajectory has a nearly vanishing
first Fourier mode. Plotted as the function of the in-\slice\ time
$\hat{\zeit}$ in \reffig{fig:f-ksconf}\,(f), these rapid episodes are
well resolved.

While the {\fFslice} resolves the reduced flow arbitrarily close to the
$\hat{x}_1 = 0$ \sliceBord, by sampling it with the in-slice time, this
symmetry reduction scheme works only as long as the amplitude of the
first mode is nonzero. For turbulent flows the {\fFslice}\ appears
empirically valid for regions of dynamical interest; in all our numerical
simulations of long-time ergodic trajectories of \KS\ system (as well as
of \NSe\rf{WiShCv14}) we have never encountered exactly vanishing first mode.

In summary, we recommend that the `{\fFslice}', a very simple symmetry
reduction prescription  \refeq{eq:FixFirstMode}, easily implemented
numerically, be used to reduce the \Un{1}\ or \SOn{2}\ symmetry of spatially
extended systems, such as shear flows in periodic domains. For example,
Avila \etal\rf{AvMeRoHo13} have recently shown that localized \rpo s have
features strikingly similar to turbulent puffs. The {\fFslice}
visualisations of the \statesp, such as \reffig{f:ks}\,(b), should help
illuminate details of the role such solutions play in transition to
turbulence.

\acknowledgements

We are indebted to
Xiong Ding,
Ashley P. Willis
and
Francesco Fedele
for stimulating discussions, to
Daniel Borrero-Echeverry
and Hugues Chat\'{e} for
a critical reading of the manuscript,
and to the anonymous referee, whose suggestions had led to many
improvements to this letter.
P.~C.\ thanks the family of late G.~Robinson,~Jr.\
and
NSF~DMS-1211827 for support.
Matplotlib library\rf{Hunter2007} was used to produce the figures in this
letter. The Matlab code used in our computations is
available on a GitHub repository\rf{ffmSliceGitHub}.

 \bibliography{../../bibtex/siminos}
\end{document}